\documentclass[aps,prl,reprint,showpacs,superscriptaddress]{revtex4-1}

\usepackage{amsmath,amssymb,times,bm}
\usepackage[pdftex]{graphicx,color}
\usepackage[pdftex,bookmarks=false,pdfborder={0 0 0},colorlinks=true,linkcolor=magenta,citecolor=cyan,urlcolor=cyan]{hyperref}
\graphicspath{{../../figure/},{../figure/},{./}}
\newcommand{\edit}[1]{\textcolor{red}{Revision needed. \\}}
\begin{document}

\title{Spontaneous modulation of superconducting phase in Kitaev ladder}
\date{\today}
\author{Sho Nakosai}
\affiliation{Condensed Matter Theory Laboratory, RIKEN, Wako, Saitama, Japan}
\author{Yukio Tanaka}
\affiliation{Department of Applied Physics, Nagoya University, Nagoya, Aichi, Japan}
\author{Tai Kai Ng}
\affiliation{Department of Physics, Hong Kong University of Science and Technology, Clear Water Bay, Hong Kong, China}
\affiliation{Hong Kong Academy for Gifted Education, Shatin, Hong Kong, China}
\author{Naoto Nagaosa}
\affiliation{RIKEN Center for Emergent Matter Science (CEMS), Wako, Saitama, Japan}
\affiliation{Department of Applied Physics, University of Tokyo, Bunkyo, Tokyo, Japan}

\begin{abstract}
	We study theoretically the two Kitaev chains put in parallel, i.e., Kitaev ladder, coupled
	by Josephson junction. The $\pi$-junction between the Majorana bound states at the ends
	of the chains competes with the usual Josephson coupling along the chain, and this frustration
	leads to the modulation of the phase difference of the superconducting order parameter
	between the two chains. We show that this modulation gives the double degeneracy of the
	ground states, which can be manipulated by external electric and magnetic fields.
\end{abstract}
\maketitle
\textit{Introduction.--}
The Kitaev chain~\cite{PhysUsp.44.131}, one-dimensional (single channel) $p$-wave superconducting wire, is
well known as a prototype of topological superconductors~\cite{RevModPhys.83.1057,JPSJ.81.011013,RepProgPhys.75.076501,AnnuRevConMatPhys.4.113}.
The prominent feature is that it hosts Majorana modes as topologically protected states at the ends of the
system. They have zero energy, which are energetically separated from other states
by the superconducting gap, and then can remain localized at the ends.
Since Majorana modes carry half the degrees of freedom of ordinary fermion, two
Majorana modes sitting on the two ends constitute one fermionic mode. The occupied and
unoccupied states of the mode have the same energy, and the ground state is doubly
degenerate.
As the consequence of peculiar properties of Majorana states, two Josephson-coupled topological
superconductors produce a fermionic Andreev bound state with energy $\epsilon (\phi)$ which
has $4 \pi$ periodicity~\cite{EurPhysJB.37.349,PhysRevB.79.161408} in sharp contrast to $2\pi$
periodicity in usual Josephson junctions.
This is due to the conservation of the fermion parity, i.e., the energy levels of
states with even and odd fermion parity cross at $\phi=2\pi$ and they come back to the
initial states only when the phase runs additional $2\pi$ if there is no parity-mixing
process.

Majorana states exhibit non-Abelian statistics~\cite{PhysRevB.61.10267,RevModPhys.80.1083,NatPhys.7.412}
and, therefore, have potential application in topological quantum
computation~\cite{AnnPhys.303.2}. The implementation have been done in some experimental
setups: (i) spin-orbit-coupled quantum nanowire proximitized to $s$-wave
superconductors~\cite{PhysRevB.79.094504,PhysRevLett.105.077001,PhysRevLett.105.177002,Science.336.1003,NatPhys.8.795,
NatPhys.8.887,NanoLett.12.6414} and the
recent progress in Ref.~\cite{NatNano.10.232}; (ii) magnetic adatoms on $s$-wave
superconductors~\cite{PhysRevB.84.195442,PhysRevB.88.020407,Science.346.602,arXiv1505.06078}; and (iii)
induced superconductivity in two-dimensional topological
insulators~\cite{PhysRevLett.100.096407,NatPhys.10.638,NatNanotech.10.593,PhysRevLett.109.056803,PhysRevLett.114.066801}.
Let us consider the two Kitaev chains aligned in parallel where
a Majorana state sit on each end of each chain.
If they are totally decoupled, Majorana fermions remain zero energy states, therefore the ground
state degeneracy is 4.
However, in general, a Majorana state sitting on an end can couple with other Majorana states
on the same end of the neighboring chains.
This lifts the degeneracy and the energy of the Majorana states become nonzero by the amount of
the tunneling element.

\begin{figure}[b]
	\centering
	\includegraphics[width=.55\hsize]{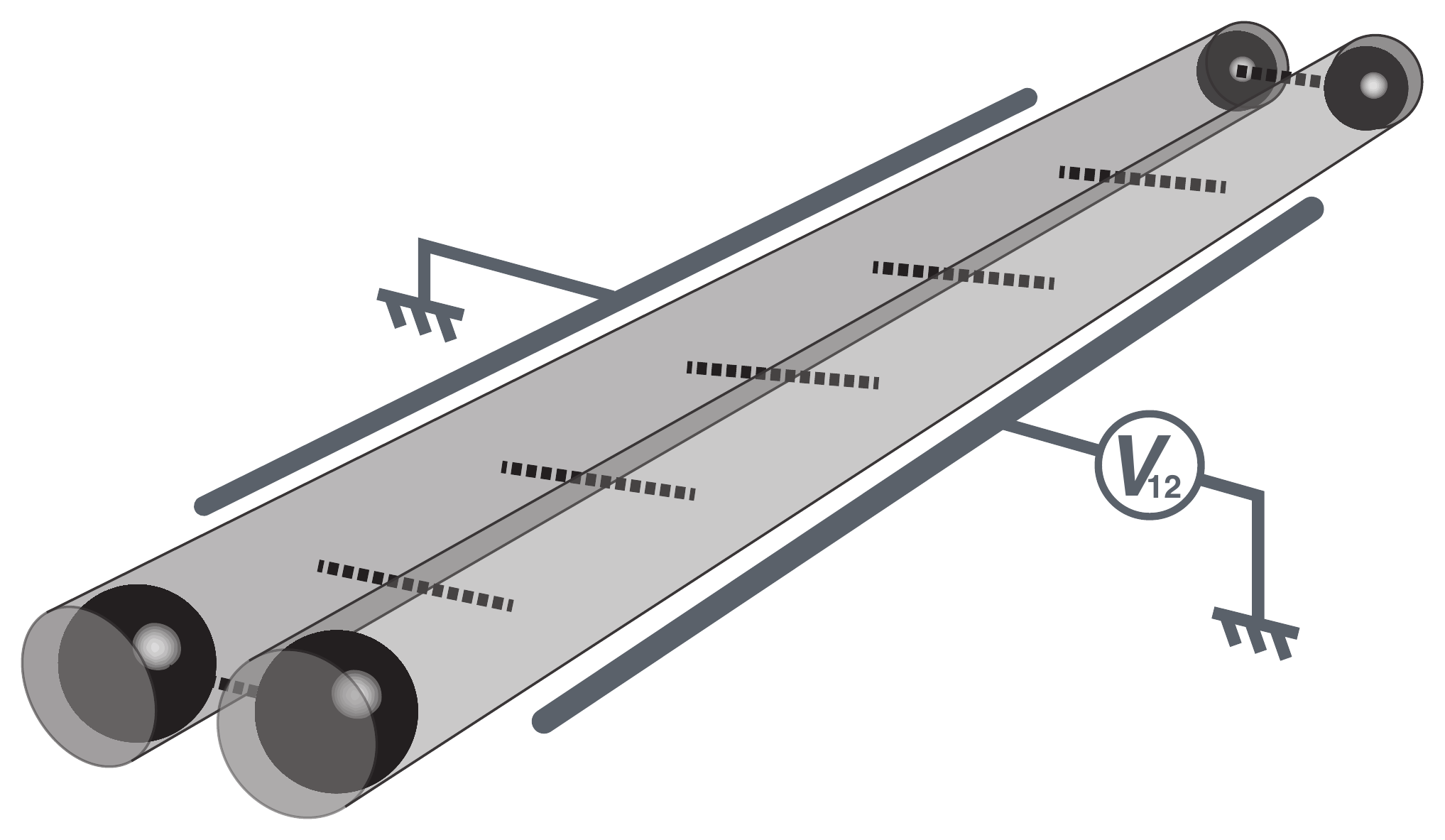}
	\caption{\label{fig:image}Schematic picture of our model representing coupled
	two Kitaev chains which 	are connected through the single-electron tunneling shown
	in dotted lines. The solid spheres show
	Majorana states localized at the ends of the chains. $V_{12}$ denotes the gate
	voltage relatively applied to chains $1$ and $2$.
}
\end{figure}

In this paper, we theoretically study two parallel Kitaev chains, i.e., Kitaev ladder shown
in Fig.~\ref{fig:image},
and discuss the effect of the inter-chain Josephson coupling.
The competing Josephson couplings, i.e., that between Majorana bound states and that between
bulk states, bring about the modulation in the phase of the superconducting order parameter and
double degeneracy of the ground states in contrast to the naive expectation
that all the degeneracies are lifted.
One can also find some interesting works on aligned Kitaev chains in
Refs.~\cite{PhysRevLett.109.156803,PhysRevLett.110.117002,PhysRevB.89.174514,PhysRevB.90.045118,
	PhysRevB.91.214507,PhysRevLett.111.116402,1509.05345}.
The change in the Majorana edge mode for the two-dimensional array of Kitaev chains due to
the spontaneous phase modulation was also discussed in Ref.~\cite{NewJPhys.15.085002}.
In contrast to these earlier works,
we focus here the double degeneracy and its manipulation in the Kitaev ladder.

\begin{figure}
	\centering
	\includegraphics[width=.8\hsize]{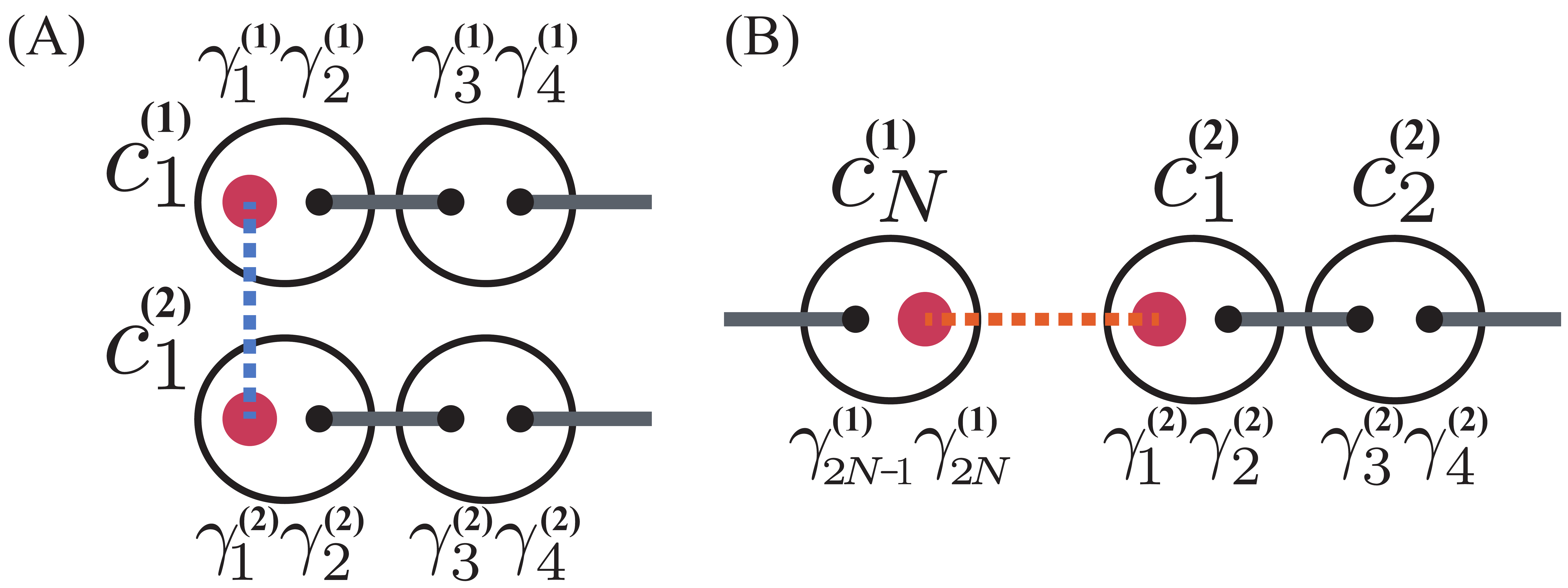}
	\caption{\label{fig:coupling}(Color online) Two configurations of the coupling between Majorana edge states;
	parallel alignment (A) and head-to-tail configuration (B).
	The solid circles represent complex fermions which are decomposed into two Majorana states
	shown in black dots. The isolated Majorana states are colored with red.
}
\end{figure}

\textit{Majorana bound states.--}
Figure~\ref{fig:image} shows the system we are considering,
which is the two superconducting chains with a finite
tunneling between them. Particularly, our interest is in the case when these chains
are topologically nontrivial phases. The topological feature is manifested in the
emergence of Majorana states at the boundaries of the
system depicted by solid spheres in Fig.1. The Kitaev chain model provides
the simplest description of this physics, which is a one-dimensional $p$-wave
superconducting system~\cite{PhysUsp.44.131}.
The Hamiltonian for two Kitaev chains with tunneling connecting them reads
\begin{align}
	H =
	\sum_{\alpha=(1),(2)}\sum_{j=1}^{N}
	\left[ \left( -t c_j^{\alpha\dagger} c_{j+1}^\alpha
		+ \Delta_\alpha c_j^\alpha c_{j+1}^\alpha
		+ \mathrm{h.c.} \right) \right.
		\nonumber \\
	\left. - \mu \left( c_j^{\alpha\dagger} c_j^{\alpha} - \frac{1}{2} \right) \right]
	+ \sum_{j=1}^{N} \left( \lambda c_j^{(1)\dagger} c_j^{(2)}
	+ \mathrm{h.c.} \right),
	\label{eq:h}
\end{align}
where $t$ is the nearest neighbor hopping strength, $\mu$ is the chemical potential,
$\Delta_\alpha=|\Delta|e^{i\theta_\alpha}$ is the $p$-wave
superconducting order parameter with the phase $\theta_\alpha$, and $\lambda$
is the coupling between these two chains.
We put parentheses for the indices indicating chains to avoid confusions.
The electron operators $c_j^\alpha$ can be decomposed into
Majorana operators $\gamma_i^\alpha$ in the following way
\begin{align}
	\gamma_{2j-1}^\alpha &=
	\exp \left( i \frac{\theta_\alpha}{2} \right) c_j^\alpha
	+ \exp \left( -i \frac{\theta_\alpha}{2} \right) c_j^{\alpha\dagger},
	\nonumber \\
	\gamma_{2j}^{\alpha} &=
	- i \exp \left( i \frac{\theta_\alpha}{2} \right) c_j^{\alpha}
	+ i \exp \left( -i \frac{\theta_\alpha}{2} \right) c_j^{\alpha\dagger}.
	\label{eq:transf}
\end{align}
First, let us consider decoupled chains by setting $\lambda=0$, and for simplicity, we choose particular
set of parameters, $|\Delta|=t>0$ and $\mu=0$, where the system is in a topological phase.
The Hamiltonian is written by using Majorana operators as
\begin{align}
	H_0 = i t \sum_{\alpha}\sum_{j=1}^{N-1} \gamma_{2j}^\alpha \gamma_{2j+1}^\alpha.
	\label{eq:dimer}
\end{align}
In the ground state, $\gamma_{2j}^\alpha \gamma_{2j+1}^\alpha$ form dimmers
depicted by the gray bonds in Fig.~\ref{fig:coupling}.
$\gamma_1^\alpha$ and $\gamma_{2N}^\alpha$ are isolated Majorana modes
which do not appear in the Hamiltonian,
and thus these operators are related to zero energy states~\cite{PhysUsp.44.131}.
In general cases with different parameters, the corresponding states cannot be represented
only by such completely isolated Majorana modes, but they still have zero energy and localized
at the ends of the system until the gap in the bulk spectrum is closed
at $|\mu|=2|t|$. Their wavefunctions decay exponentially with the length scale of
$\xi_\mathrm{M}\sim t/|\Delta|$ (the lattice constant is set to be a unit).

Now, we turn on a finite single electron tunneling $\lambda$.
The tunneling term in Eq.~\eqref{eq:h} can be written with respects to
Majorana operators as
\begin{align}
	H_\mathrm{int} &{} = \sum_{i} \lambda ( c_i^{(1)\dagger} c_i^{(2)} + c_i^{(2)\dagger} c_i^{(1)})
	\nonumber \\
	&{} = \frac{i\lambda}{2} \sum_{i} \left\{
		\sin\left( \frac{\theta_1-\theta_2}{2} \right)
		\left( \gamma_{2i-1}^{(1)} \gamma_{2i-1}^{(2)} + \gamma_{2i}^{(1)}
		\gamma_{2i}^{(2)} \right) \right.
		\nonumber \\
		& \hspace{2em} \left.
		+ \cos\left( \frac{\theta_1-\theta_2}{2} \right)
		\left( \gamma_{2i-1}^{(1)} \gamma_{2i}^{(2)} - \gamma_{2i}^{(1)}
		\gamma_{2i-1}^{(2)} \right)
	\right\}.
	\label{eq:t}
\end{align}
Here, we considered the interchain tunneling process only between the
parallel sites. We will treat it as a perturbation.
In the dimer limit (Eq.~\eqref{eq:dimer}),
the effective Hamiltonian can be easily derived.

Let us start with $\gamma_n^{(1)} \gamma_m^{(2)}$ with
$n,m \ne 1, 2N$, i.e., the Majorana fermions
participating in the dimer formation.
For this coupling, there is a finite gap in the unperturbed
Hamiltonian, and hence the second order perturbation
theory can be applied. A straightforward calculation gives
the perturbation energy $-(\lambda^2/16t) \cos(\theta_1-\theta_2)$
except a constant energy shift.
This dependence on the phase difference is expected for the ordinary Josephson
coupling in superconducting junctions.

For $\gamma_n^{(1,2)}$ with $n=1,2N$, there is no unperturbed
Hamiltonian, and hence the degenerate perturbation theory must be
applied.
Picking up terms which includes only these Majorana fermions in
Eq.~\eqref{eq:t} are
\begin{align}
	& \frac{i\lambda}{2}
	\sin\left( \frac{\theta_1-\theta_2}{2} \right)
	\left( \gamma_{1}^{(1)} \gamma_{1}^{(2)} + \gamma_{2N}^{(1)}
	\gamma_{2N}^{(2)} \right)
	\nonumber \\
	& = \lambda \sin\left( \frac{\theta_1-\theta_2}{2} \right)
	(a^\dagger a + b^\dagger b -1 )
	\label{eq:MF}
\end{align}
where we have introduced (complex) fermion operators as
$a = (\gamma^{(1)}_1 +i \gamma^{(2)}_1)/2$,
$b = (\gamma^{(1)}_{2N} +i \gamma^{(2)}_{2N})/2$.
Therefore, the ground state energy becomes
$- \lambda |\sin\left( \frac{\theta_1-\theta_2}{2} \right)|$
by appropriate choice of
$|0\rangle_i$ and $|1\rangle_i$ labeled by $i=a,\,b$ satisfying $a | 0 \rangle_a = 0$ and
$| 1 \rangle_a = a^{\dagger} | 0 \rangle_a$ (and the same for $b$).

Of crucial importance is the dependence on the phase difference, which is the sinusoidal form.
This should be contrasted to the anomalous Josephson coupling of Majorana states
discussed in, for example,
Refs.~\cite{EurPhysJB.37.349,PhysRevB.79.161408}. The configuration there is depicted in
Fig.~\ref{fig:coupling}~(B), and the coupling arises from the tunneling term
$(\lambda c_N^{(1)\dagger} c_1^{(2)} + \mathrm{h.c.})$ which results in
$i\lambda \cos \left(\left( \theta_1 - \theta_2 \right)/2\right) \gamma_{2N}^{(1)} \gamma_1^{(2)}$.
One can understand this form by looking at the last term in Eq.~\eqref{eq:t}.
The characteristic $4\pi$ periodicity is originating from the factor $1/2$ in the $\cos$ function.

\begin{figure}
	\centering
	\includegraphics[width=.85\hsize]{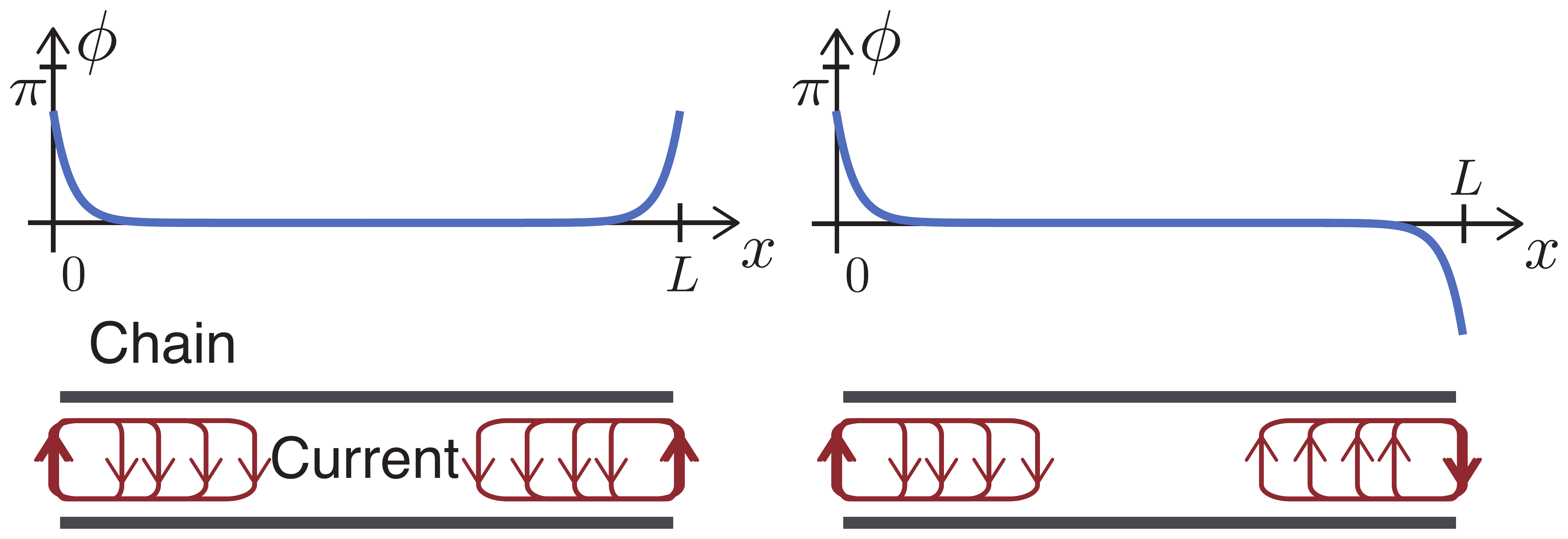}
	\caption{\label{fig:sc} (Color online) Schematic depiction of kink-antikink (left) and
	kink-kink (right) configurations of the phase difference and emerging
	supercurrent flowing in the system due to the modulation in the phase. The kink-antikink
configuration has lower energy.}
\end{figure}

\textit{GL theory of the superconducting phases.---}
We consider the GL theory of the phases of the superconducting order parameter
based on the above result.
In a system with an infinite length, the Hamiltonian is written as
\begin{align}
	H & = \int_0^L dx \left\{
		\rho_{\mathrm{s}} \left[ {(\partial_x \theta_1)}^2
		+ {(\partial_x \theta_2)}^2 \right]
	- J \cos\left( \theta_1 - \theta_2 \right) \right\}
  \nonumber \\
	& \hspace{.2\hsize}- \lambda \left| \sin\left( \frac{\theta_1-\theta_2}{2} \right) \right|_{x=0,L}
	\label{eq:jj}
\end{align}
where $J =\lambda^2/16t$ and $\rho_{\mathrm{s}}$ is the
superfluidity density, i.e., the stiffness of the phase.
We introduce the center of mass phase $\theta$ and the phase difference
$\phi$ as $\theta_1 = \theta + \phi/2$ and
$\theta_2 = \theta - \phi/2$. The solutions can be obtained by the ordinary stationary
equation for both fields
$\theta$ and $\phi$. Since it is obvious that $\theta$ takes a constant value,
we set it to be $0$. As for the
phase difference $\phi$, we obtain the sine-Gordon equation
\begin{align}
	\partial_x ^2 \phi - J \sin \phi = 0,
	\label{eq:sg}
\end{align}
The boundary term in Eq.~\eqref{eq:jj} will induce the deviation from the trivial solution $\phi=0$,
which can be treated by the part of the solution to Eq.~\eqref{eq:sg}
as given by
\begin{align}
	\phi(x) = \pm 4 \tan^{-1} \left[ e^{(x-x_0)/\xi_{\phi}} \right]
	\label{eq:kink}
\end{align}
with the assumption that $\phi(x\to-\infty) = 0$, $\phi(x\to\infty) = \pm 2\pi$,
and $(d/dx) \phi(x\to\pm\infty) = 0$. This solution has a kink at $x=x_0$
which is fixed by the competition between the bulk term
and boundary term in Eq.~\eqref{eq:jj}.
The slope of the kink is determined by the typical length scale
$\xi_\phi=\sqrt{\rho_{\mathrm{s}}/J}$.
The boundary condition at the ends of the system becomes
\begin{align}
	{\left[ \partial_x \phi \pm \lambda \cos \frac{\phi}{2} \right]}_{x=0,\,L} = 0.
	\label{eq:bc}
\end{align}
arising from the last term in Eq.~\eqref{eq:jj}.
The point is that the phase difference has a finite value at the vicinity of the ends to
minimize the energy due to the Majorana Josephson coupling whereas the phase difference
in the bulk is close to $0$ as in a usual Josephson junction.
We emphasize that this spatial modulation of the phases along the chains originates
from the zero energy Majorana bound states arising as topologically protected states.
The position of the kink in the solutions Eq.~\eqref{eq:kink}, i.e. $x_0$,
is pinned by this boundary condition to a particular value with which
\begin{align}
	\phi(x=0) = \pm 4\tan^{-1} \left( \sqrt{1+\frac{16J}{\lambda^2}}
	- \frac{4\sqrt{J}}{\lambda} \right)
	\label{eq:phi0}
\end{align}
is satisfied.
Eventually, the solutions to Eq.~\eqref{eq:sg} with open boundaries are obtained as
\begin{align}
	{\phi(x)} \!=\! \pm 4
	\! \left\{ \tan^{\!-\!1} \!\left[ e^{-(x-x_0)/\xi_\phi} \right]\! \pm \tan^{\!-\!1}
	\!\left[ e^{(x-x_0-L)/\xi_\phi} \right] \right\},
	\label{eq:sol}
\end{align}
where the ones with the plus sign in the curly bracket have the same sign at the two ends
(see the left panel in Fig.~\ref{fig:sc}) and the ones with the minus sign have the opposite
signs (the right panel). We call them as kink-antikink
and kink-kink solutions, respectively. Here we assume that the solutions
are approximated simply by superpositions of
two kink solutions.
We evaluate the corresponding energies by substituting above solutions into Eq.~\eqref{eq:jj}.
It turns out that the kink-antikink
solutions result in lower energy. The gap in energy depends on the stiffness of the superconducting
order parameter and
decreases exponentially as $\exp\left( -L/\xi_\phi \right)$.
The consequent supercurrent by the spatial modulation of the phase is calculated by using
the expression $j=\delta H/\delta \phi$
with given configurations Eq.~\eqref{eq:sol} as~\footnote{The explicit form of Eq.~\eqref{eq:lce} with
$\rho_\mathrm{s}$ is $j_{\mathrm{end}} = \pm \left( \frac{8 J^{\frac{3}{2}} \lambda \sqrt{\rho_\mathrm{s}}}{16J
	+ \lambda^2 \rho_\mathrm{s}} + \frac{2 \sqrt{J}\, \lambda }{\sqrt{16J + \lambda^2 \rho_\mathrm{s}}} \right)$.}
\begin{align}
	j_{\mathrm{bulk}}(x) = &{}
	\mp  2J\, \mathrm{sech}^{-1} \left[ f_{J,\lambda}(x) \right] \tanh
	\left[ f_{J,\lambda}(x) \right],
	\label{eq:lcb}
	\\
	j_{\mathrm{end}} = &{}
	\pm \left( \frac{8 J^{\frac{3}{2}} \lambda}{16J + \lambda^2}
	+ \frac{2 \sqrt{J}\, \lambda }{\sqrt{16J + \lambda^2}} \right),
	\label{eq:lce}
\end{align}
where
\begin{align}
	f_{J,\lambda}(x) = \sqrt{J} x - \sinh^{-1} \left( \frac{4\sqrt{J}}{\lambda} \right).
\end{align}
The distribution of the current is schematically depicted in the second row in Fig.~\ref{fig:sc}.
One can see that it can be regarded as the loop current on each half of the system.
The amount of the current through the bulk Josephson coupling is compensated by
those at the ends through the Majorana bound states.
It is naturally expected that these loop current generate magnetic field.
We simply approximate the induced field as magnetic
dipoles sitting on the ends. The interaction between the dipoles are antiferromagnetic,
and clearly seen from Fig.~\ref{fig:sc},
it further reduces the energy of a kink-antikink configuration.
In summary, Majorana Josephson coupling produces non-zero phase difference
near the ends of the system with plus or
minus sign resulting in a kink-kink or kink-antikink configuration. The kink-antikink
configuration is energetically preferable.

\textit{Low energy physics.---}
Finally, we construct an effective theory describing the low energy physics
in our model based on the above discussion.
We introduce the following set of Pauli matrices; $\sigma_{i}^{z}$ corresponding
to the occupancies of $a$ and $b$ fermions and $\tau_{i}^{z}$
the sign of the phase difference $\phi$ at the left and right ends.
The Majorana Josephson coupling is represented by
\begin{align}
	H_m = m (\sigma_a^z \tau_a^z + \sigma_b^z \tau_b^z),
	\label{eq:hmass}
\end{align}
where we set $m=\lambda \sin\left( |\phi(0)|/2 \right)$ (see Eq.~\eqref{eq:phi0}).
The interaction between kink configuration at the both ends can be embedded in
\begin{align}
	H_r = - r \tau_a^z \tau_b^z,
	\label{eq:hkk}
\end{align}
where $r$ is a positive value since the kink-antikink configuration is energetically
favored as discussed (note that such a
configuration corresponds to the matrix element with the same eigenvalues of $\tau_{i}^{z}$).
$r$ scales as $\exp\left( -L/\xi_\phi \right)$.

Another possible term comes from the overlap of the Majorana bound states at
two ends of each chain.
It is described by using complex-fermion operators $a$ and $b$
\begin{align}
	H_w &{} = w (i \gamma_{1}^{(1)} \gamma_{2N}^{(1)} +
	i \gamma_{1}^{(2)} \gamma_{2N}^{(2)} ) \nonumber \\
	&{} = w (a^\dagger b - b^\dagger a),
	\label{eq:w}
\end{align}
where $w\propto \exp\left( -L/\xi_\mathrm{M} \right)$.
Equation~\eqref{eq:w} can be rewritten in terms of $\sigma_{i}^{x,y}$ as
\begin{align}
	H_w = w ( \sigma_a^x \sigma_b^y - \sigma_a^y \sigma_b^x ).
\end{align}

\begin{figure}
	\centering
	\includegraphics[width=\hsize]{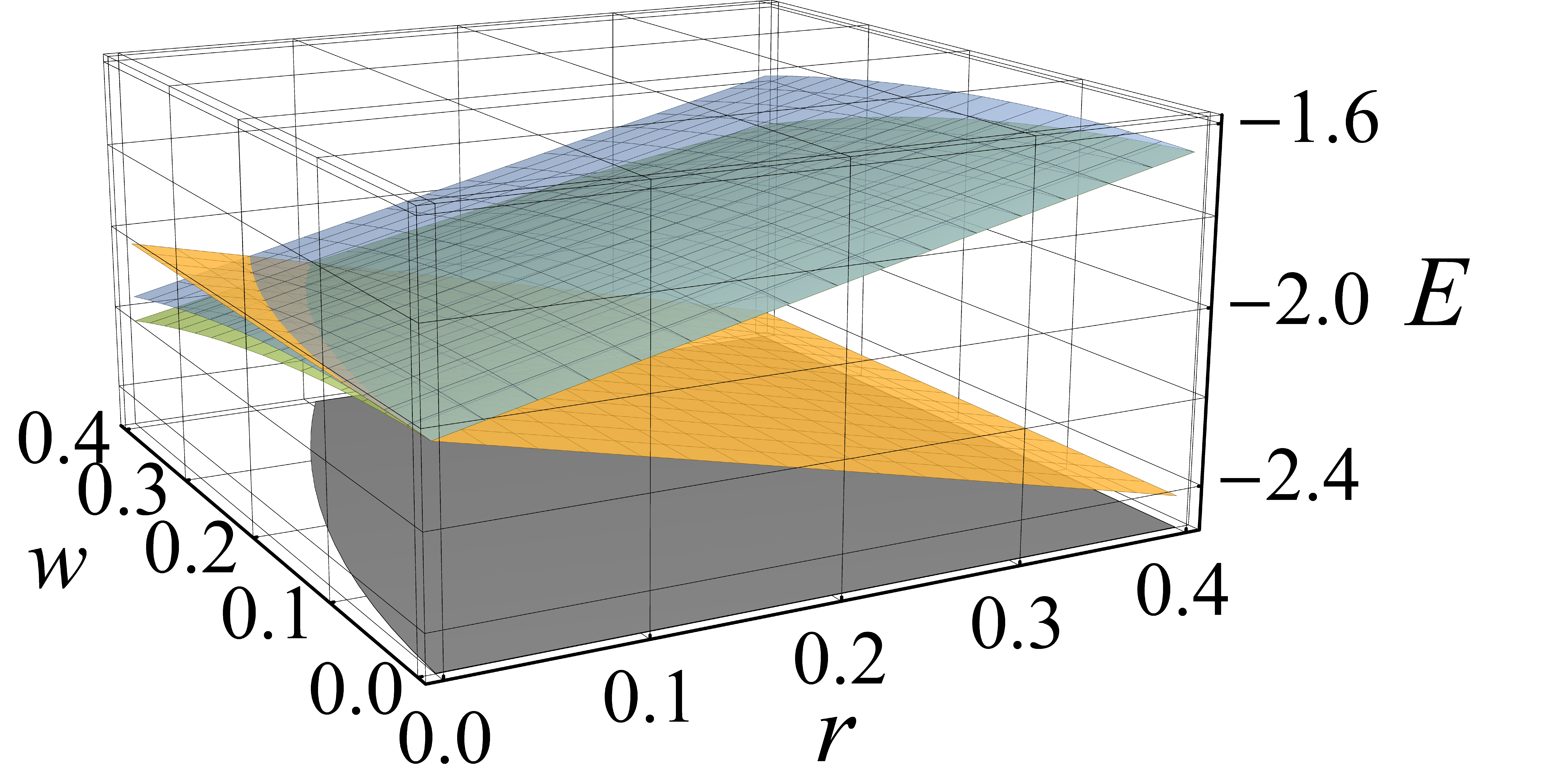}
	\caption{\label{fig:e} (Color online) Eigenvalues of $H_\mathrm{eff}$ as functions of $r$ and $w$.
	$m=1$ and $|\Gamma|=0.2$ are chosen. The yellow surface shows doubly degenerate eigenvalue.
	The shaded region on the bottom represents the area in the parameter space where this doubly degenerate
ground state is realized.}
\end{figure}

We also consider an additional non-trivial term expressing the macroscopic quantum tunneling
with the matrix element $\Gamma$ as
\begin{align}
	H_{\mathrm{MQT}} = -\Gamma (\tau_1^+ + \tau_2^+)
	- \Gamma^* (\tau_1^- + \tau_2^-).
	\label{eq:mqt}
\end{align}
It describes quantum tunneling process connecting kink- and antikink-configurations.
The total effective Hamiltonian $H_\mathrm{eff}=H_m+H_r+H_w+H_\mathrm{MQT}$
is in the form of $16\times16$ matrix and the energy eigenvalues can be obtained by
the diagonalization. Here $m$ is considered to dominate other parameters
since both $r$ and $w$ decay exponentially in
the length of the system.
Figure~\ref{fig:e} shows four low-lying eigenvalues out of 16 eigenvalues of $H_\mathrm{eff}$.
Here we set $m=1.0$ and $|\Gamma|=0.2$.
The shaded area shown in the bottom of the graph shows the set of parameters $r$ and $\Gamma$
with which the doubly degenerate eigenvalues
(shown in the yellow curve) have the lowest values, i.e., the ground state is degenerate.
Note that $r$ and $w$ have different length scales $\xi_\phi$ and $\xi_\mathrm{M}$, then,
we expect that in a system with $\xi_\phi>\xi_\mathrm{M}$ the magnitude of $r$ exceeds that of $w$ and
the ground state is degenerate.
For example, in the experiment found in Ref.~\cite{Science.346.602} they reported the localization length
of the Majorana states are comparable to the size of an adatom on the substrate
superconductor~\cite{PhysRevLett.114.106801}.
The states corresponding to the degenerate energy are ferromagnetic in pseudospin of fermionic states
$\langle \sigma_{1}^{z} \rangle = \langle \sigma_{2}^{z} \rangle$ and their configurations of the superconducting
order parameter are superposition of two kink-antikink configurations.
We can argue that by taking into account the phase degrees of freedom and by the virtue of
anomalous Majorana Josephson coupling the ground state degeneracy holds even when the finite
size effect and the quantum tunneling processes play roles.

\textit{Conclusion and outlook.--}
We investigate the role of modulation in the superconducting order
parameter in two Kitaev chains put in parallel.
When those are in the topological phase, single electron tunneling between Majorana states sitting
at the ends is allowed and gives anomalous Josephson coupling, which results in spatially modulated
superconducting phase. Due to this modulation, the ground state are doubly degenerate
even though we take into
account the overlap of Majorana wavefunctions.
This double degeneracy can act as a qubit, and its manipulation is an important issue.
As discussed above,
the kink-antikink solutions don't have the net magnetic field,
and hence cannot be biased by the uniform external magnetic field.
However the local magnetic field applied only on one end of the system lifts
the degeneracy.

The other means to control the state is to put the potential difference $V_{12}(t)$
between the two chains, which modulates the phase of the tunneling matrix element
$\Gamma$ in Eq.~(\ref{eq:mqt}) between the two degenerate configurations
of the superconducting order parameter.
This comes from the canonical conjugate relation between the
charge difference $Q_1 - Q_2$ between the two chains and
the phase difference $\phi$ of the superconducting order parameter.
$V_{12}(t)$ is coupled to $Q_1 - Q_2$. After integrating over
$Q_1 - Q_2$, we obtain the term $i V_{12} d \phi/dt $ in the Lagrangian,
which put the phase factor $\exp{ \left( ic V_{12} \int_0^L  d x \bar{\phi}(x) \right) }$
to $\Gamma$ ($c$: a constant depending on the capacity of the system, $\bar{\phi}(x)$:
the difference of $\phi(x)$ between the two stable kink configurations).
Combining these two external fields, one can manipulate
the dynamics of the qubit.

\textit{Acknowledgments.--}
The authors thank Patrick~A.~Lee for the fruitful discussion.
SN was supported by Grant-in-Aid for JSPS Fellows. TKN acknowledges support by
HKRGC through grant No.~HKUST3/CRF/13G.
This work was supported by Grant-in-Aid for Research
Activity Start-up (Grant No.~15H06858);
Grant-in-Aid for Scientific Research (S) (Grant No.~24224009); and
Grant-in-Aid for Innovative Areas ``Topological Materials Science''
(Grant No.~15H05853) from MEXT, Japan.

\end{document}